\documentclass[3p,times,procedia]{elsarticle}
\usepackage{nupha_ecrc}

\volume{00}
\firstpage{1}
\journalname{Nuclear Physics A}
\runauth{Frithjof Karsch et al.}
\jid{nupha}
\jnltitlelogo{Nuclear Physics A}
\usepackage{amssymb}
\usepackage[figuresright]{rotating}
\usepackage{graphicx}
\newcommand{\SB}{S_{\hspace{-2pt}B}}
\newcommand{\SP}{S_{\hspace{-2pt}P}}

\begin{document}
\begin{frontmatter}

\dochead{}

\title{Conserved Charge Fluctuations from Lattice QCD and the 
Beam Energy Scan}

\author{F. Karsch$^{1,2}$, A. Bazavov$^3$, H.-T. Ding$^4$, P. Hegde$^4$, 
O. Kaczmarek$^2$, E. Laermann$^2$, Swagato Mukherjee$^1$, H. Ohno$^{1,5}$, 
P. Petreczky$^1$, C. Schmidt$^2$, S. Sharma$^1$, \\
W. Soeldner$^6$, P. Steinbrecher$^2$, M. Wagner$^7$}
\address{$^1$Physics Department, Brookhaven National Laboratory, Upton, NY 11973, USA\\
$^2$Fakult\"at f\"ur Physik, Universit\"at Bielefeld, D-33615 Bielefeld,
Germany\\
$^3$Department of Physics and Astronomy, University of Iowa,
Iowa City, Iowa 52240, USA\\
$^4$Key Laboratory of Quark \& Lepton Physics (MOE) and Institute
of Particle Physics, Central China Normal University, \\
Wuhan 430079, China\\
$^5$Center for Computational Sciences, University of Tsukuba,
Tsukuba, Ibaraki 305-8577, Japan\\
$^6$Institut f\"ur Theoretische Physik, Universit\"at Regensburg,
D-93040 Regensburg, Germany\\[-2pt]
$^7$Physics Department, Indiana University, Bloomington, IN 47405,
USA
}

\begin{abstract}
We discuss the next-to-leading order Taylor expansion of ratios of 
cumulants of net-baryon number fluctuations. We focus on the relation
between the skewness ratio, $\SB\sigma_B\equiv\chi_3^B/\chi_1^B$, 
and the kurtosis ratio, $\kappa_B\sigma_B^2\equiv\chi_4^B/\chi_2^B$.
We show that differences in these two cumulant ratios are small for
small values of the baryon chemical potential. The next-to-leading
order correction to $\kappa_B\sigma_B^2$ however is approximately three 
times larger than that for $\SB\sigma_B$.
The former thus drops much more rapidly
with increasing beam energy, $\sqrt{s_{NN}}$. 
We argue that these generic patterns are consistent with current data on
cumulants of net-proton number fluctuations measured by the STAR
Collaboration at $\sqrt{s_{NN}}\ge 19.6$~GeV.
  \end{abstract}


\end{frontmatter}

\section{Introduction}

Fluctuations of 
conserved charges of strong interactions have long been considered as a 
set of sensitive observables to explore the structure of the phase
diagram of Quantum Chromodynamics (QCD). They currently are the 
most promising experimental observables in the search for a possible
critical point in the phase diagram of QCD that gets 
performed with the beam energy scan (BES) program at the Relativistic Heavy 
Ion Collider (RHIC). Although the results on net-electric charge
\cite{Adamczyk:2014fia,Adare:2015aqk} and net-proton number \cite{STARp08,STARp20} 
fluctuations obtained from the first BES runs performed at 
RHIC did not yet provide clear cut evidence for the existence of a critical point,
the collected data on charge fluctuations show an intriguing dependence on
the beam energy which at present is poorly understood even qualitatively.
The published data on cumulants of net-proton number fluctuations
\cite{STARp08} and, in particular, the still preliminary data set,
which covers a larger transverse momentum range 
\cite{STARp20}, show obvious deviations from the thermodynamics of a 
hadron resonance gas (HRG), e.g. the ratios of even as well as odd order 
cumulants differ from unity and different mixed ratios,
formed from even and odd order cumulants, are not identical. This may not be too 
surprising as HRG model calculations are not expected to give an accurate 
description of higher order cumulants close to the QCD (phase) transition 
region.  It, however, raises the question whether the observed pattern seen
in measured net-proton fluctuations can be
understood in terms of QCD thermodynamics or whether other effects are
responsible for this. 

We will discuss net-proton number 
fluctuations at small values of the baryon chemical potential, $\mu_B$. 
In particular, we will point out basic features observed in  
ratios of cumulants of net-proton number fluctuations in the BES at RHIC, 
which can not be accommodated in HRG model calculations, but are naturally 
explained in QCD using a next-to-leading (NLO) order expansion of
cumulant ratios in terms of $\mu_B$.

\section{Cumulants of net-baryon number fluctuations}
We will discuss here the structure of ratios of net-baryon number cumulants
that can be formed with the first four cumulants which are related to
mean ($M_B$), variance ($\sigma_B^2$), skewness ($\SB$) and kurtosis
($\kappa_B$) of net baryon number distributions, 
\begin{eqnarray}
R_{12}^B(T,\mu_B,\mu_Q,\mu_S)&\equiv& \frac{\chi_1^B(T,\mu_B,\mu_Q,\mu_S)}{\chi_2^B(T,\mu_B,\mu_Q,\mu_S)} 
\equiv \frac{M_B}{\sigma_B^2} \;\; , \;\; 
R_{31}^B(T,\mu_B,\mu_Q,\mu_S)\equiv \frac{\chi_3^B(T,\mu_B,\mu_Q,\mu_S)}{\chi_1^B(T,\mu_B,\mu_Q,\mu_S)} 
\equiv \frac{\SB\sigma_B^3}{M_B} \;\; , \nonumber \\
R_{42}^B(T,\mu_B,\mu_Q,\mu_S)&\equiv& \frac{\chi_4^B(T,\mu_B,\mu_Q,\mu_S)}{\chi_2^B(T,\mu_B,\mu_Q,\mu_S)} 
\equiv \kappa_B \sigma_B^2 \;\; .
\label{cumulants}
\end{eqnarray}
Here the $n$-th order cumulants are obtained
as partial derivatives of the QCD pressure, $P(T,\mu_B,\mu_Q,\mu_S)$,
with respect to the chemical potentials,
\begin{equation}
\chi_n^B (T,\mu_B,\mu_Q,\mu_S) = 
\frac{\partial^n P/T^4}{\partial(\mu_B/T)^n} \;\;\;\; {\rm or}\;\;\;\; 
\chi_{nm}^{BX} (T,\mu_B,\mu_Q,\mu_S) = 
\frac{\partial^{(n+m)} P/T^4}{\partial(\mu_B/T)^n \partial(\mu_X/T)^m} 
\;\; ,\;\; X=Q,\ S \;\;  .
\label{chiB}
\end{equation}
Using Eq.~\ref{cumulants} we also obtain
$R_{32}^B\equiv \SB \sigma_B$, 
as $R_{32}^B=R_{31}^B R_{12}^B$. 
For this 
ratio and those 
introduced in Eq.~\ref{cumulants} 
we may set up
Taylor expansions in terms of the chemical potentials. Here 
it is convenient to introduce constraints that resemble thermal 
conditions met in heavy ion collisions, i.e. we demand strangeness
neutrality $M_S\equiv \chi_1^S(T,\mu_B,\mu_Q,\mu_S)=0$ and 
fix a relation between net-electric charge ($M_Q$) and net baryon 
number ($M_B$). We may choose $M_Q=r M_B$ with $r=0.4$, which resembles
electric charge to baryon number ratio in the incident beams in heavy ion 
collisions at RHIC and LHC.
With these conditions the chemical potentials $\mu_Q$ and $\mu_S$ become
functions of $\mu_B$, i.e. to leading order one has 
$\mu_S/\mu_B=s_1(r)$, $\mu_Q/\mu_B=q_1(r)$ \cite{freeze}.
The ratios $R_{nm}^B$ introduced in 
Eq.~\ref{cumulants} then become functions of $T$ and $\mu_B$ only. 
They may then be Taylor expanded in $\mu_B$,
\begin{equation}
R_{nm}^B(T, \mu_B) = R_{nm}^{B,0} (T)  +  R_{nm}^{B,2}(T) \left( \frac{\mu_B}{T}\right)^2 +{\cal O}(\mu_B^4) \;\; .
\label{Taylor}
\end{equation}
We will focus here on the relation between the expansion coefficients
for the skewness ratio $R_{31}^B\equiv \SB\sigma^3_B/M_B$ and those of
the kurtosis ratio $R_{42}^B\equiv \kappa_B \sigma_B^2$. In a medium with
vanishing strangeness and electric charge chemical potential
this relation is, in fact, quite simple:

\begin{equation}
\hspace*{-3.0cm}\underline{\mu_Q=\mu_S=0:} \hspace*{4.0cm}
R_{42}^{B,0}(T) = R_{31}^{B,0}(T) \;\; , \;\; R_{42}^{B,2}(T) = 3 R_{31}^{B,2}(T) 
\label{LO}
\end{equation}
However, when implementing the constraints $M_S=0$ and $M_Q/M_B=r$, such
simple relations no longer hold. In fact, in this case the leading order (LO) 
coefficients are related through
\begin{equation}
\hspace*{-1.5cm}\underline{M_S=0,\ M_Q/M_B=r:} \hspace*{1.0cm}
R_{42}^{B,0}(T) - R_{31}^{B,0}(T) =
\frac{s_1 (\chi_{31}^{BS}\chi_2^B - \chi_{11}^{BS}\chi_4^B) + 
q_1 (\chi_{31}^{BQ}\chi_2^B-\chi_{11}^{BQ}\chi_4^B)}{
(\chi_2^B + s_1 \chi_{11}^{BS} + q_1 \chi_{11}^{BQ})\chi_2^B}  \;\; ,
\end{equation}
The corresponding expressions for the NLO 
corrections are somewhat more involved. They will be presented elsewhere
\cite{later}.

\begin{figure}[t]
\begin{center}
\includegraphics[width=73mm]{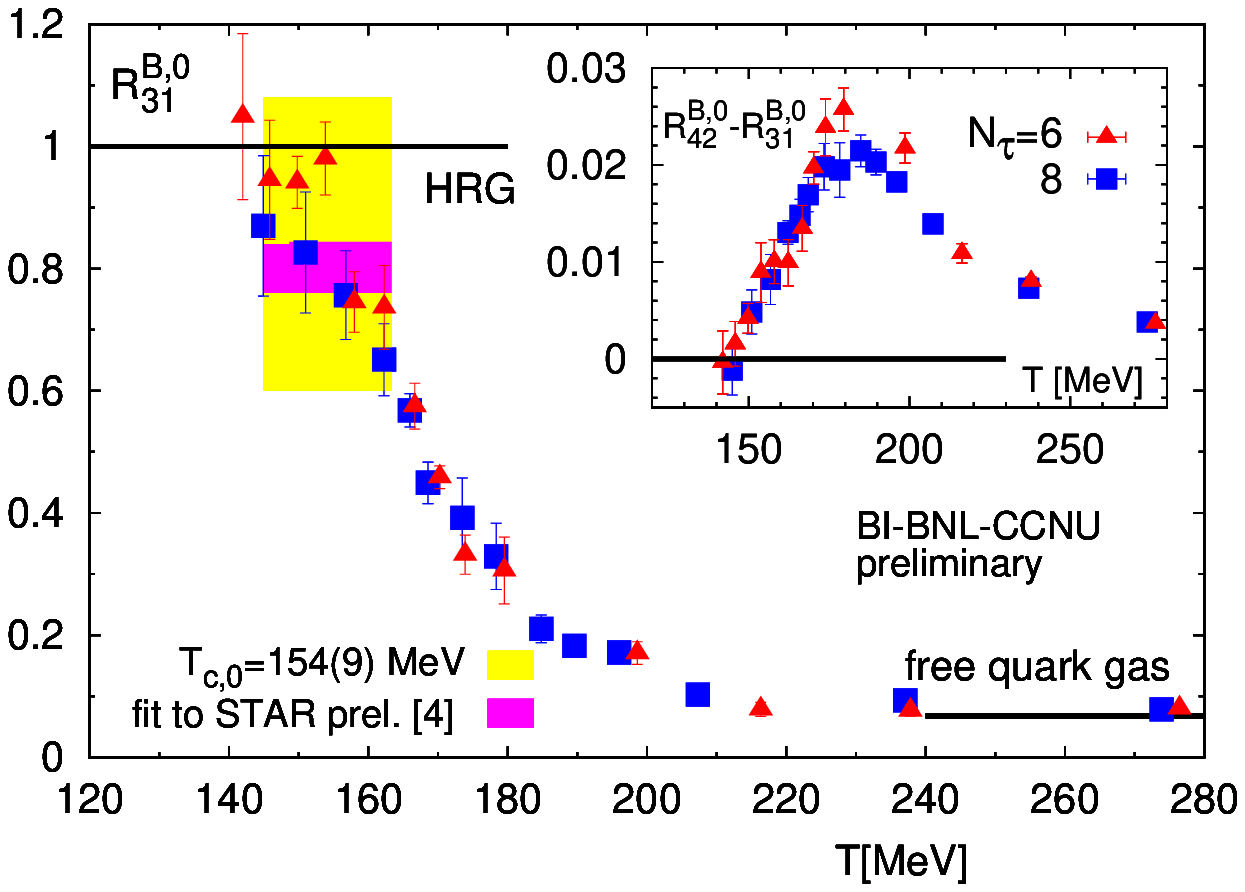}
\includegraphics[width=73mm]{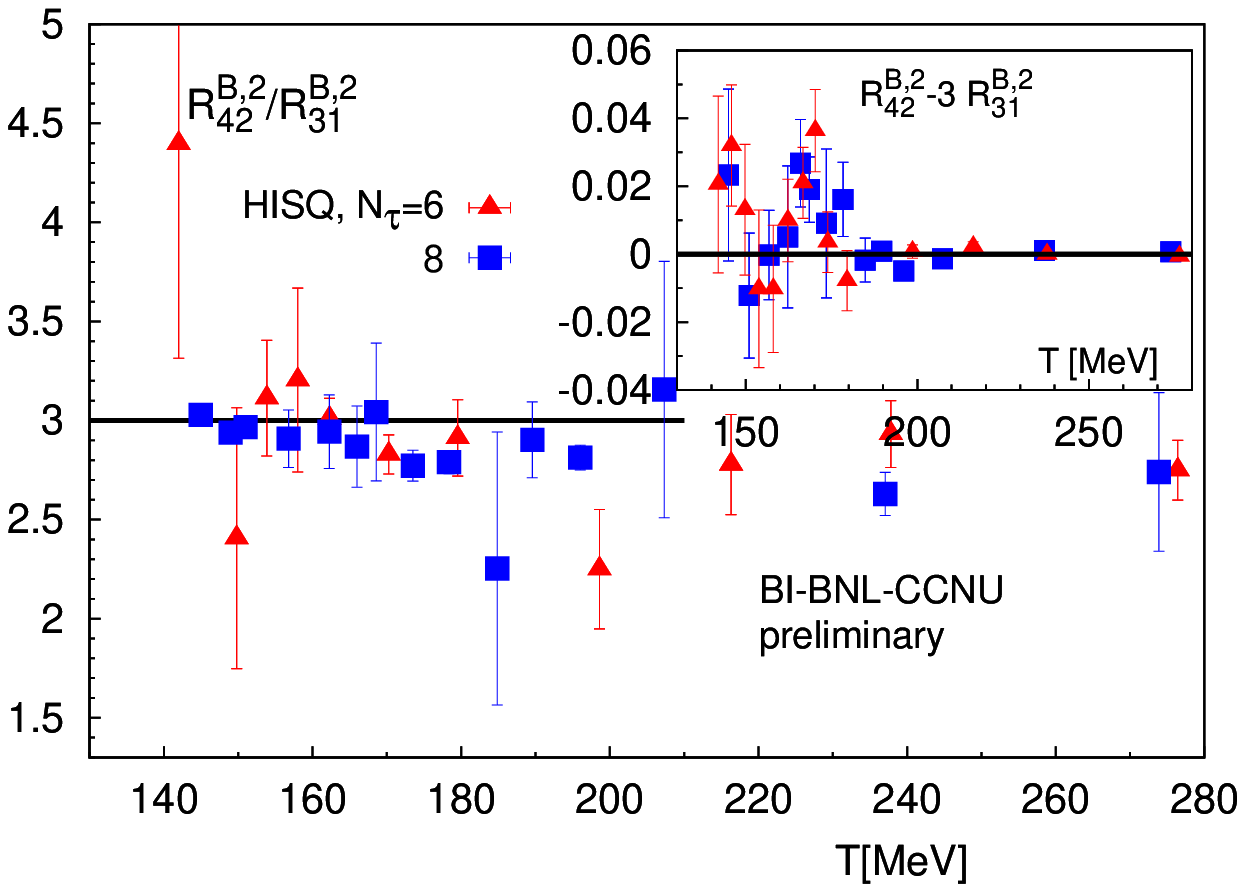}
\caption{Left: The leading order ($R_{31}^{B,0}$) result for the 
skewness ratio $R_{31}^B$
versus temperature calculated in (2+1)-flavor QCD on lattices of size
$(4N_\tau)^3\cdot N_\tau$ with $N_\tau =6,\ 8$ for a strangeness neutral
system with $r=0.4$. 
The insertion shows the difference between the LO expansion 
coefficients for $R_{42}^B$ and $R_{31}^B$. Right: The ratio of the NLO 
expansion coefficients, $R_{42}^{B,2}$ and $R_{31}^{B,2}$.  
The insertion shows the corresponding difference $R_{42}^{B,2} -3R_{31}^{B,2}$. 
Horizontal lines in the right hand figure 
corresponds to the result for $\mu_S=\mu_Q=0$. For details see discussion
in text.
}
\label{fig:R42-31}
\end{center}
\end{figure}

In Fig.~\ref{fig:R42-31}(left) we show results for $R_{31}^{B,0}$, i.e. 
the leading order 
result for $R_{31}^B$ at $\mu_B=0$. The insertion in this figure shows the
difference $R_{42}^{B,0}(T) - R_{31}^{B,0}(T)$. As can be seen, for 
temperatures in the crossover region, $T_{c,0}=(154\pm 9)$~MeV (yellow band in main panel), the magnitude 
of this difference is at most $2\%$ of 
$R_{31}^{B,0}(T)$ but may reach about $10\%$ at $T\simeq 180$~MeV. This 
suggests that the skewness ratio $R_{31}^B=\SB\sigma_B^3/M_B$ and the
kurtosis ratios $R_{42}^B=\kappa_B\sigma_B^2$ should be almost identical
at the highest RHIC energies, where $\mu_B/T\simeq 0.15$.

In Fig.~\ref{fig:R42-31}(right) we show the ratio of the NLO expansion
coefficients $R_{42}^{B,2}(T)$ and $R_{31}^{B,2}(T)$. For $\mu_Q=\mu_S=0$
it is straightforward to show that this ratio equals three as stated in
Eq.~\ref{LO}. In the constraint case this, however, does not need to be the 
case. In fact, in the infinite temperature limit the ratio varies between
$5/3$ and $2$, depending on the value of $M_Q/M_B=r$. Precise calculations of 
the ratio $R_{42}^{B,2}/R_{31}^{B,2}$ are demanding as one needs to 
evaluate $6^{th}$ order cumulants and both expansion coefficients may
change sign in the temperature range of interest. With the presently 
available statistics this causes the large errors on  
$R_{42}^{B,2}/R_{31}^{B,2}$ 
seen in Fig.~\ref{fig:R42-31}(right) for some values of $T$. 
It is, however, evident that the 
ratio of NLO expansion coefficients stays close to $3$ in a wide 
$T$-range. Hence one may expect that the dependence of the
kurtosis ratio, $\kappa_B\sigma_B^2=\chi_4^B/\chi_2^B$, on $\mu_B$, 
and thus on $\sqrt{s_{NN}}$,
is significantly larger than that of the skewness
ratio $\SB\sigma_B^3/M_B= \chi_3^B/\chi_1^B$.

\section{Cumulants of net-proton number fluctuations}
\begin{figure}[t]
\begin{center}
\includegraphics[width=68mm]{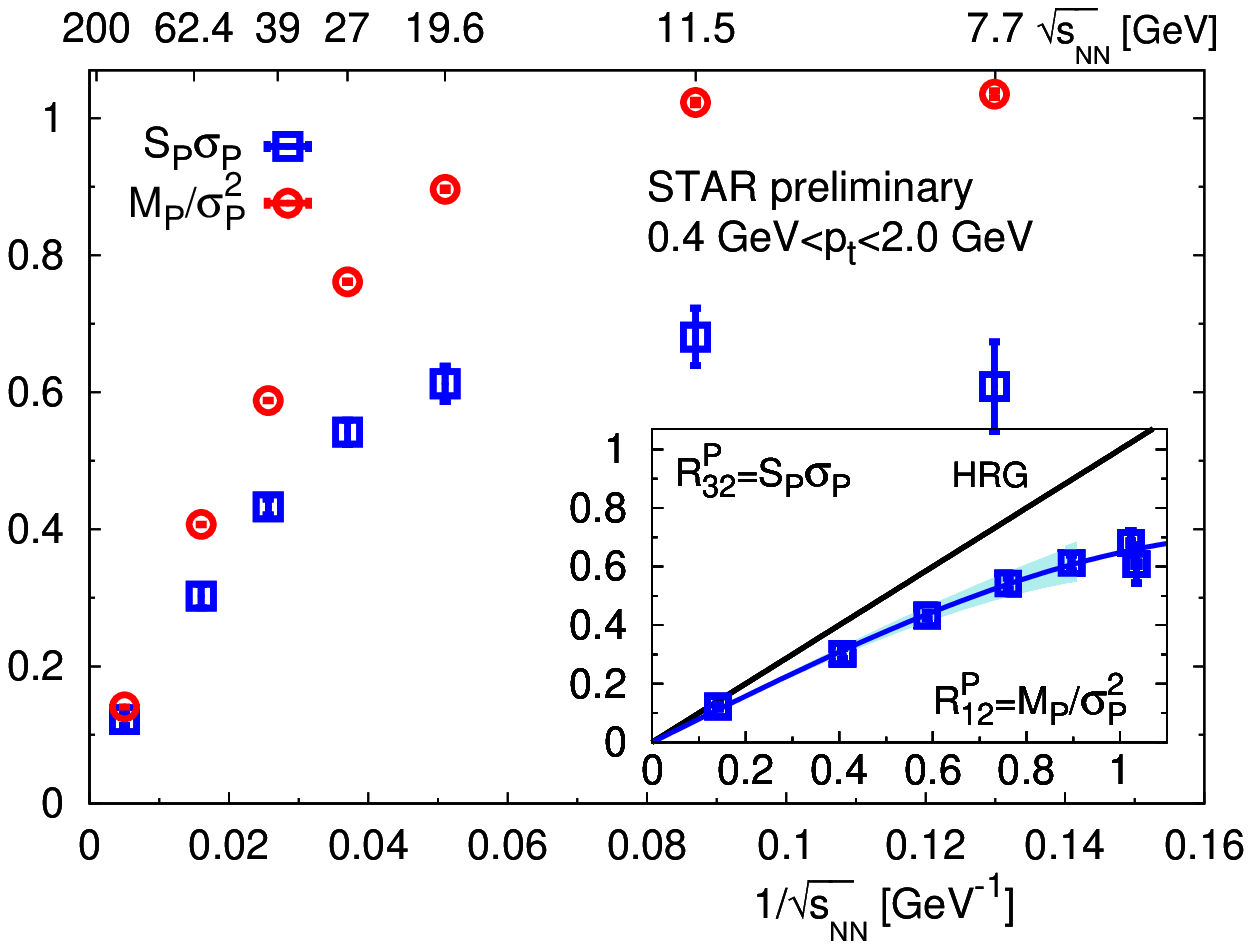}            
\includegraphics[width=68mm,height=51mm]{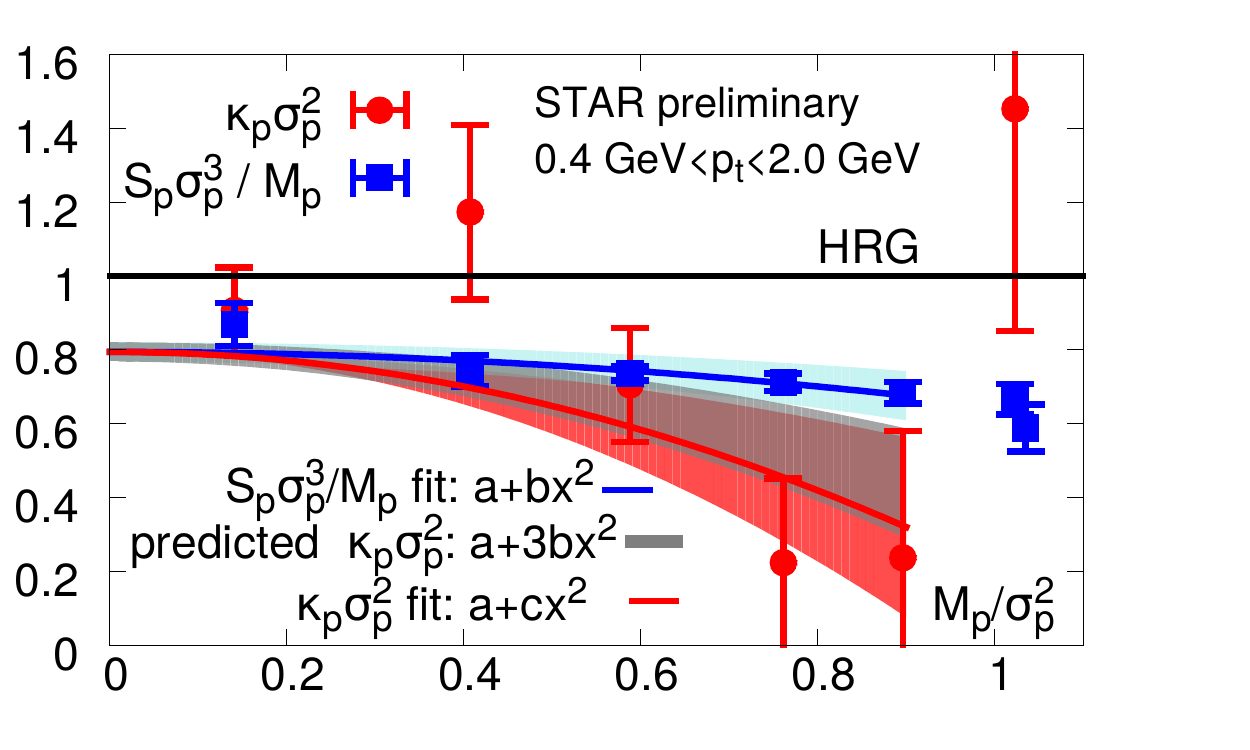}
\caption{Left: Preliminary data for $R_{12}^P=M_P/\sigma_P^2$ and 
$R_{32}^P=\SP \sigma_P$ obtained by the STAR Collaboration in
the transverse momentum interval, $0.5~{\rm GeV}\le p_t\le 2.0~{\rm GeV}$.
Right. Results for $R_{31}^P=\chi_3^P/\chi_1^P$ constructed from the 
data shown in the left hand part of the figure and preliminary data
for $R_{42}^P=\chi_4^P/\chi_2^P$ (the data point for $R_{42}^P$ at $\sqrt{s_{NN}}=7.7$~GeV is not shown). For details see discussion in the text.
}
\label{fig:skewness}
\end{center}
\end{figure}

We compare these generic features of the relation of LO and NLO 
expansion coefficients of the Taylor expansion of cumulant ratios
$R_{42}^B$ and $R_{31}^B$ with experimental data on ratios of
cumulants of net proton number fluctuations measured in the BES 
at RHIC. To do so, we also note that we may eliminate
the dependence of these cumulant ratios on $\mu_B$, by solving the 
Taylor series expansion for another ratio, e.g. 
$R_{12}^B(T,\mu_B)=M_B/\sigma_B^2\equiv R_{12}^{B,1}(T)\mu_B/T
+{\cal O}(\mu_B^3)$. Eliminating $\mu_B/T$ in Taylor expansions
in favor of $R_{12}^B$ is possible as long as the relation between both
is unique. This will not be the case in general, but seems to hold in the
$(T,\mu_B)$ regime covered in the BES. As can be seen from
Fig.~\ref{fig:skewness}(left), along the freeze-out line the ratio $R_{12}^P$ 
is a monotonically decreasing function of the beam energy, i.e. $R_{12}^P$
rises with increasing $\mu_B$. 

In Eq.~\ref{Taylor} we thus may replace $\mu_B/T$ by 
$ \left( R_{12}^{B,1}\right)^{-1} M_B/\sigma_B^2$. 
Experimentally one cannot directly measure net-baryon number fluctuations
and their cumulants. One rather has access only to cumulants of 
net-proton number fluctuations. It then is 
appropriate to consider the ratios $R_{42}^P$ and $R_{31}^P$
as functions of $R_{12}^P$ and test to what extent the generic 
features discussed for the corresponding ratios $R_{42}^B$ and $R_{31}^B$
are reflected in the data.  In Fig.~\ref{fig:skewness}
we show preliminary data on ratios of various net-proton number
cumulants obtained by the STAR Collaboration in the transverse 
momentum range $0.4 {\rm GeV} \le p_t\le 2.0 {\rm GeV}$. The left
hand figure shows that $\SP\sigma_P < M_P/\sigma_P^2$ holds in the entire 
energy range covered in the BES at RHIC. This is clearly
different from HRG model predictions, where these two
cumulant ratios are identical. Such a difference, however, naturally 
arises in QCD thermodynamics. As can be seen in Fig.~\ref{fig:R42-31}(left) 
the ratio $R_{31}^B = R_{32}^B/R_{12}^B$ is smaller than unity for all $T$.
A quadratic fit to the STAR data for  $R_{31}^P$ vs. $R_{12}^P$,
for $R_{12}^P\le 0.9$, 
or equivalently for
beam energies
$\sqrt{s_{NN}}\ge 19.6$~GeV, yields $R_{31}^P= 0.80(4)-0.15(5) (R_{12}^P)^2$.
The intercept at $R_{12}^P=0$, i.e. at
$\mu_B=0$, is shown
in Fig.~\ref{fig:R42-31}(left) as a horizontal bar. It is 
consistent with a freeze-out temperature at or below the QCD transition 
temperature, $T_{c,0}$. This also is consistent with
freeze-out temperatures obtained from an analysis of cumulants of
net-electric charge fluctuations as functions of $R_{12}^P$ 
\cite{Bazavov:2015zja}.

Fig.~\ref{fig:skewness}(right) shows the STAR data 
for $R_{31}^P$ and $R_{42}^P$. The latter have large statistical and
systematic errors. The grey band shown in this figure is the expected
behavior of  $R_{42}^P$ when using knowledge on $R_{31}^P$  as input 
and assuming that the data on net-proton number fluctuations follow the 
generic behavior discussed above for net-baryon number fluctuations
in QCD at small values of $\mu_B$, i.e. it shows three times the slope
obtained from a fit to $R_{31}^P$ (light blue band). Performing a fit to the
data for $R_{42}^P$ that is constrained at $\mu_B=0$ by assuming
$R_{42}^P (\mu_B=0) = R_{31}^P (\mu_B=0)$ yields, 
$R_{42}^P = 0.80 -0.59(30) (R_{12}^P)^2$. This is shown by the light red
band. Although 
errors on the data are 
still large, this result is consistent with an expected 
factor three larger curvature coefficient for the data on $R_{42}^P$
with respect to the data on $R_{31}^P$. 

\vspace*{-1pt}
\section{Conclusions}
Data on cumulants of higher order net-proton 
number fluctuations taken at RHIC at beam energies $\sqrt{s_{NN}}\ge 19.6$~GeV
are consistent with expectations deduced from QCD calculations for cumulants of net-baryon number fluctuations performed in
a NLO Taylor expansion in $\mu_B$. In particular, the
strong decrease of $\kappa_P\sigma_P^2$ relative to the mild 
variation of $\SP\sigma_P$ is consistent with ''non-critical''
behavior of cumulant ratios.

\vspace*{1pt}
\noindent
{\it Acknowledgements:}
This work has been partially supported through an ALCC grant of the 
U.S. Department of Science, the DOE under Contract No. DE-SC0012704,
the BMBF of Germany under grant no. 05P15PBCAA and the research fund 
No. 11450110399 from the National Science Foundation of China.

\bibliographystyle{elsarticle-num}

\end{document}